\documentclass[aps,pra, twocolumn, showpacs,amsfonts,amsmath,floatfix, superscriptaddress,10pt, nofootinbib]{revtex4-2}
\pdfoutput=1
\usepackage{asymptote}
\usepackage{amsmath}
\usepackage{mathtools}
\usepackage{amsfonts}
\usepackage{amssymb}
\usepackage{amsthm}
\usepackage{stmaryrd}
\usepackage{physics}
\usepackage{bbm}
\usepackage[dvipsnames]{xcolor}
\usepackage[colorlinks=true,linkcolor=teal,citecolor=Maroon,urlcolor=Maroon]{hyperref}
\usepackage[capitalize]{cleveref}

\newcommand{\secref}[1]{Section~\hyperref[#1]{\ref*{#1}}}
\newcommand{\appref}[1]{Appendix~\hyperref[#1]{\ref*{#1}}}
\newcommand{\tabref}[1]{Table~\hyperref[#1]{\ref*{#1}}}
\newcommand{\figref}[1]{Fig.~\hyperref[#1]{\ref*{#1}}}
\newcommand{\sfigref}[2]{Fig.~\hyperref[#1]{\ref*{#1}(#2)}}
\renewcommand{\vec}[1]{\mathbf{#1}}

\DeclareMathOperator{\IM}{im}
\DeclareMathOperator{\SPAN}{span}
\DeclareMathOperator{\image}{im}

\newcommand{\prlsection}[1]{\textit{#1.---}\hspace{+0em}}

\begin{document}
\title{Single-shot and measurement-based quantum error correction via fault complexes}

\author{Timo Hillmann}
\email{timo.hillmann@rwth-aachen.de}
\affiliation{Xanadu Quantum Technologies Inc., Toronto, Ontario, M5G 2C8, Canada}
\affiliation{Department of Microtechnology and Nanoscience (MC2), Chalmers University of Technology, SE-412 96 Gothenburg, Sweden}

\author{Guillaume Dauphinais}
\affiliation{Xanadu, Toronto, Ontario, M5G 2C8, Canada}

\author{Ilan Tzitrin}
\email{ilan@xanadu.ai}
\affiliation{Xanadu, Toronto, Ontario, M5G 2C8, Canada}

\author{Michael Vasmer}
\affiliation{Xanadu, Toronto, Ontario, M5G 2C8, Canada}
\affiliation{Perimeter Institute for Theoretical Physics, Waterloo, Ontario, N2L 2Y5, Canada}
\affiliation{Institute for Quantum Computing, Waterloo, Ontario, N2L 3G1, Canada}

\date{\today}

\begin{abstract} 
Photonics provides a viable path to a scalable fault-tolerant quantum computer. The natural framework for this platform is measurement-based quantum computation, where fault-tolerant graph states supersede traditional quantum error-correcting codes. However, the existing formalism for foliation---the construction of fault-tolerant graph states---does not reveal how certain properties, such as single-shot error correction, manifest in the measurement-based setting. We introduce the fault complex, a representation of dynamic quantum error correction protocols particularly well-suited to describe foliation. Our approach enables precise computation of fault tolerance properties of foliated codes and provides insights into circuit-based quantum computation. Analyzing the fault complex yields improved thresholds for three- and four-dimensional toric codes, a generalization of stability experiments, and the existence of single-shot lattice surgery with higher-dimensional topological codes.
\end{abstract}

\maketitle

\prlsection{Introduction}
Photonic platforms for quantum computing~\cite{bourassa_blueprint_2021, tzitrin_fault-tolerant_2021, bartolucci_fusion-based_2023, alexander2025manufacturable, walshe_linear-optical_2025, aghaee_rad_scaling_2025} are well-suited to measurement-based quantum computing (MBQC)~\cite{briegel_measurement-based_2009}.
In this paradigm, in contrast to circuit-based quantum computation (CBQC), the central object is not a quantum error-correcting (QEC) code but a fault-tolerant graph state (FTGS). Various methods for constructing FTGS exist~\cite{raussendorf_fault-tolerant_2007,nickerson_measurement_2018,newman_generating_2020}, with the concept of foliation~\cite{bolt_foliated_2016, brown_universal_2020} offering a prescription for any Calderbank-Shor-Steane (CSS) code~\cite{steane_error_1996, calderbank_good_1996}.
Here, we introduce the \emph{fault complex}, a representation of faults in a dynamic QEC protocol (rather than a static QEC code) formulated in the language of homology and chain complexes~\cite{bombin_homological_2007, breuckmann_quantum_2021,hastings_weight_2017, evra_decodable_2022}.
We focus on fault complexes obtained via foliation, which we recast in the language of homology as a tensor product~\cite{tillich_quantum_2014,zeng_higher-dimensional_2019} between a CSS code and a repetition code.
This formalism allows us to easily calculate properties like the fault distance, and also applies to CBQC where it represents repeated rounds of error correction.
Analysis of the fault complex enables us to understand the decoding of single-shot codes in MBQC and achieve improved error thresholds for three-dimensional (3D) and 4D toric codes~\cite{dennis_topological_2002, vasmer_three-dimensional_2019, quintavalle_single-shot_2021, higgott_improved_2023}.
Through the explicit calculation of the homology groups of the fault complex, we generalize the notion of stability experiments~\cite{gidney_stability_2022}, which in turn allows us to conclude that single-shot lattice surgery is possible in higher-dimensional topological codes.
The fault complex provides a formal language for fault-tolerant protocols, similar to how homological descriptions serve CSS codes, laying the groundwork for future advances in fault-tolerant quantum computation.

\prlsection{Background and Notation}
An $[n, k, d]$ binary linear code $C$ forms a $k$-dimensional subspace within the $n$-dimensional vector space over $\mathbb{F}_2$.
Such a code can be defined by a parity-check matrix (PCM) $H$, where the codewords are the elements of $\ker H$.
The distance $d$ is the minimum Hamming weight of any nonzero codeword.
We use $e_i$ to denote the $i^\text{th}$ unit vector and $[M]_j$ to denote the $j^\text{th}$ row of a matrix $M$.
We write $\vec 0$ and $\vec 1$ for the all-zero and all-one vectors, respectively.

A stabilizer code~\cite{gottesman_stabilizer_1997} is the quantum analogue of a linear code and is defined by an abelian subgroup, $\mathcal S$, of the Pauli group with $-I\notin\mathcal S$.
The codespace is the $+1$ eigenspace of $\mathcal S$; logical operators commute with $\mathcal S$ but are not themselves in $\mathcal S$.
The notation $\llbracket n, k, d \rrbracket$ describes a code in which the stabilizer group is generated by $m = n - k$ independent generators, which are themselves elements of the $n$-qubit Pauli group $\mathcal{P}_n$.
The number of encoded qubits is $k$ and the distance $d$ is determined by the minimum weight non-trivial logical operator.
A stabilizer code admitting a set of stabilizer generators that are each either $X$-type or $Z$-type operators is referred to as a CSS code~\cite{steane_error_1996, calderbank_good_1996}.
These codes can be described by two classical binary linear codes
with PCMs $H_X$ and $H_Z$ representing stabilizer generators as tensor products of $X$ and $I$, and $Z$ and $I$, respectively.
Commutativity of $X$- and $Z$-type generators is expressed through the condition $H_Z^T H_X = 0$.
Error correction proceeds via the measurement of the stabilizer generators, yielding the syndrome $s$, which is the list of stabilizer eigenvalues.
This allows the detection and correction of Pauli errors that anticommute with stabilizer generators; a classical decoder is used to infer recovery operators given a syndrome.

Here, a chain complex of length $n$ denotes a collection of $\mathbb{F}_2$-vector spaces $C_i$ and linear maps $\partial_i^{C}$ called boundary operators,
\begin{align}
    \mathcal{C} = \{ 0 \} \overset{\partial_{n+1}^{C}}{\longrightarrow} C_n \stackrel{\partial_n^{C}}{\longrightarrow} \dots \stackrel{\partial_2^{C}}{\longrightarrow} C_1 \stackrel{\partial_1^{C}}{\longrightarrow} C_0 \overset{\partial_0^{C}}{\longrightarrow} \{ 0 \}, 
\end{align}
with the composition fulfilling $\partial_i^{C} \partial_{i+1}^{C} = 0$. 
We suppress superscripts if no distinction is necessary.
The quotient
$H_i(\mathcal{C}) \coloneqq \ker \partial_i / \image \partial_{i+1}$ is called the $i^\text{th}$ homology group of $\mathcal{C}$.
Associated with $\mathcal{C}$ is also a cochain complex, with coboundary operators $\delta^{i}: C_i \rightarrow C_{i+1}$ defined as $\delta^{i} = \partial_{i+1}^T$:
\begin{align}
     \mathcal{C}^T = \{ 0 \} \stackrel{\delta^{-1}}{\rightarrow} C_0 \stackrel{\delta^{0}}{\rightarrow} C_1
     \dots
     \stackrel{\delta^{n-1}}{\rightarrow} C_n
     \stackrel{\delta^{n}}{\rightarrow} \{ 0 \},
\end{align}
with cohomology groups $H^{i} \coloneqq \ker \delta^{i} / \image \delta^{i-1} $.

Each length-1 chain complex describes an $[n, k, d]$ linear code with boundary map $\partial_1 = H$ from $\mathbb{F}_2^{n}$ to the space of syndromes $\mathbb{F}_2^{r}$ with $r \geq n - k$.
Similarly, a CSS code, $\mathcal C$, can represented by a length-2 chain (sub)complex 
\begin{equation}
    \label{eq:CSS_chain_complex}
    \ldots \rightarrow C_{i+1} \stackrel{\partial_{i+1}}{\longrightarrow} C_i \stackrel{\partial_{i}}{\longrightarrow} C_{i-1} \rightarrow \ldots,
\end{equation}
where, by convention, $\partial_{i+1} = H_Z^T$ and $\partial_{i} = H_X$, such that $\partial_{i} \partial_{i+1} = 0$ is encoded in the condition that $H_X H_Z^{T} = 0$.
Identifying qubits with the space $C_i$, the code parameters are $n = \dim C_i$, $k = \dim H_i(\mathcal{C})$, and the logical operators are elements of the groups $H_i(\mathcal{C})$ and $H^i(\mathcal C)$, with the smallest weight element defining $d$.

\begin{figure*}[!ht]
    \centering
    \includegraphics{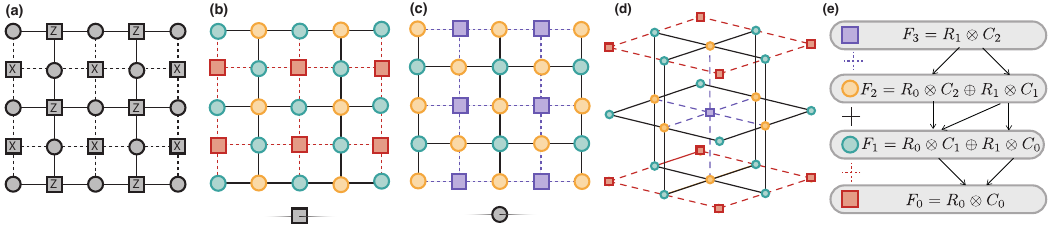}
    \caption{Foliation of the surface code viewed as a fault complex.
    \textbf{(a)} The distance-3 surface code Tanner graph with circles representing qubits, squares representing $X$ and $Z$ checks, and dashed and solid lines representing the connectivity of these checks, respectively.
    \textbf{(b)} and \textbf{(c)} Hypergraph product of (a) with a repetition code check and bit node, respectively. This yields two types of fault locations, teal and yellow, along with primal (red) and dual (purple) detectors.
    Dashed and solid lines indicate boundary maps of the fault complex. For simplicity, we omit out-of-plane connections.
    \textbf{(d)} Unit cell of the fault complex obtained by stacking alternating layers of (b) and (c). 
    A dual detector (purple) is formed by the parity of six dual fault locations (yellow). Vertical red dashed lines are omitted for clarity.
    \textbf{(e)} Mathematical structure of the fault complex and its relation to (b)-(d): colored circles and squares represent chain complex vector spaces and lines represent boundary maps.
    }
    \label{fig:surface_code_foliation}
\end{figure*}

\prlsection{Motivating example}
A common way to obtain an FTGS from a CSS code is via foliation~\cite{bolt_foliated_2016}. In this procedure, the $X$ and $Z$ Tanner graphs of the code are interpreted as alternating layers of the FTGS, and data qubits between adjacent layers are entangled with each other; see \sfigref{fig:surface_code_foliation}{a-d}. 
The computation is then performed by a sequence of adaptive local measurements on the FTGS.
We observe that foliation can be algebraically formulated as the tensor (or hypergraph~\cite{tillich_quantum_2014,zeng_higher-dimensional_2019}) product of a base CSS code with a repetition code~\footnote{We note that a similar approach has recently explored in the context of condensed matter physics~\cite{okuda_2024_anomaly,okuda_2024_anomaly2}}; see \cref{fig:surface_code_foliation}{e}.
Denote the resulting complex as $\mathcal F = \mathcal C \times \mathcal R$, where $\mathcal C$ is a length-2 chain complex describing a CSS code and $\mathcal R$ is a chain complex describing a repetition code.
The spaces of $\mathcal F$ are
\begin{equation}
    F_j 
    = \bigoplus_{\ell + m = j} R_\ell \otimes C_m 
\end{equation}
where $R_\ell$ and $C_m$ are the $\ell^\text{th}$ and $m^\text{th}$ spaces of $\mathcal R$ and $\mathcal C$, respectively.
The boundary operators of $\mathcal F$ are
\begin{align}
    \label{eq:tensor_product_boundary_maps}
    \partial_{j} =  \left( \begin{array}{c|c}
         \mathbbm{1}_{r}  \otimes \partial^C_{j}  & R \otimes \mathbbm{1}_{n_{j-1}^C}  \\ \hline
         0 & \mathbbm{1}_{c}  \otimes  \partial^C_{j-1} 
    \end{array} \right), 
\end{align}
where $\partial^C_j$ is the $j^\text{th}$ boundary operator of $\mathcal C$, and $R$ is the $r\times c$ PCM of the binary linear code.
One can observe that $\partial_2$ is the bi-adjacency matrix of the (bipartite) FTGS, whereas $\partial_1$ and $\partial_3^T$ describe the detectors (or foliated stabilizers) of the FTGS.
For example, the rows of $\partial_1$ are of the form
$(e_\alpha \otimes [H_X]_\beta | [R]_\alpha \otimes e_\beta)$,
with an analogous form for the columns of $\partial_{3}$.
These are exactly the foliated stabilizers of~\cite{bolt_foliated_2016}; see \sfigref{fig:surface_code_foliation}{d}, which illustrates a dual detector.
Each detector (row) above is triggered by a fault on one of the qubits participating in the check $[H_X]_\beta$, or by a syndrome fault in an adjacent layer. 

\prlsection{Fault complexes}
Motivated by the previous example, we define a fault complex to be a length-3 chain (sub)complex $\mathcal{F}$
\begin{align}
    \dots \rightarrow F_{i+2} \stackrel{\partial_{i + 2}}{\longrightarrow} F_{i+1} \stackrel{\partial_{i+1}}{\longrightarrow} F_{i} \stackrel{\partial_{i}}{\longrightarrow} F_{i-1} \rightarrow \dots,
\end{align}
where we define \emph{primal fault locations} to be elements of $F_{i}$ and \emph{dual fault locations} to be elements of $F_{i+1}$.
The fault complex has $n = n_{i} + n_{i+1}$ total faults, where $n_i = \dim F_i$.
The boundary map $\partial_{i+1}$ determines equivalent primal and dual faults.
We define the primal and dual \emph{detector matrices} to be $D_X = \partial_{i}$ and $D_Z = \partial_{i+2}^T$, respectively.
The syndrome of a primal fault $x \in F_{i}$ is $D_X x$ and the support of a primal detector $u \in F_{i-1}$ is $D_X^T u$, and similarly for dual faults and detectors.
In a fault complex, there is no requirement for the commutativity of the primal and dual detector matrices.
Instead, we have $\partial_{i} \partial_{i+1} = 0$, meaning that a primal fault that is equivalent to a dual fault has trivial (primal) syndrome, with an analogous interpretation for $\partial_{i+2}^T \partial_{i+1}^T = 0$.

We refer to the elements of $H^{i}(\mathcal F)$ and $H_{i+1}(\mathcal F)$ as primal and dual \emph{logical correlations}, respectively.
These represent the information that the fault complex is designed to protect.
Similarly, the elements of $H_{i}(\mathcal F)$ and $H^{i+1}(\mathcal F)$ are the primal and dual \emph{logical errors}, respectively.
These operators change the values of the logical correlations but have no syndrome and are therefore undetectable.
Note that the number of primal logical correlations (or errors) $k_i = \dim H^{i}(\mathcal F) = \dim H_{i}(\mathcal F)$ need not equal the number of dual logical correlations (or errors) $k_{i+1} = \dim H_{i+1}(\mathcal F) = \dim H^{i+1}(\mathcal F)$.
The primal (dual) \emph{fault distance} $d_i$ ($d_{i+1})$ of the fault complex is the weight of the minimal weight primal (dual) logical error.

Recently, there have been many proposals for formalizing fault-tolerant protocols~\cite{nickerson_measurement_2018,newman_generating_2020,gidney_stim_2021,derks_designing_2024,gottesman_opportunities_2022,delfosse_spacetime_2023,bombin_fault-tolerant_2023,fu_error_2024,beverland_fault_2024,li_low-density_2025}, some of which resemble our fault complexes.
In particular, our definition of a fault complex builds on the definition of a \emph{fault-tolerant cluster state} in~\cite{newman_generating_2020}.
We also note that fault complexes are distinct from the \emph{fault-tolerant complexes} of~\cite{bombin_fault-tolerant_2023}, which are defined geometrically and are limited to topological codes.
Fault complexes have an interpretation in CBQC where the faults can represent both qubit, gate, and measurement errors.
In this interpretation, $\partial_{i+1}$ is related to the gauge group of the spacetime code; see~\cite{supp} for further discussion.
Here, we concentrate on the interpretation of fault complexes in MBQC.

\prlsection{Foliated CSS codes}
We now return to our motivating example of the fault complex $\mathcal F = \mathcal C \times \mathcal R$, where $\mathcal C$ represents a CSS code and $\mathcal R$ represents a repetition code.
Suppose that $R$, $H_X$, and $H_Z$ are full rank.
From the K\"{u}nneth formula, we obtain the number of primal and dual correlations
\begin{equation*}
\begin{split}
    k_i &= \dim H_0(\mathcal R) \dim H_{i}(\mathcal C) + \dim H_1(\mathcal R) \dim H_{i-1}(\mathcal C),\\
    k_{i+1} &= \dim H_0(\mathcal R) \dim H_{i+1}(\mathcal C) + \dim H_1(\mathcal R) \dim H_{i}(\mathcal C).
\end{split}
\end{equation*}
One can show~\cite{zeng_higher-dimensional_2019,supp} that the primal and dual fault distances of $\mathcal F$ are given by
\begin{equation}
\label{eq:fault_distance}
\begin{split}
    d_{i} &= \min[d_0(\mathcal{R})d_i(\mathcal{C}), d_1(\mathcal{R})d_{i-1}(\mathcal{C})],  \\
    d_{i+1} &= \min[d_0(\mathcal{R}^T)d_{i+1}(\mathcal{C}^T), d_1(\mathcal{R}^T)d_{i}(\mathcal{C}^T)],
\end{split}
\end{equation}
where $d_i(\mathcal C)$ and $d_i(\mathcal C^T)$ are equal to the minimal weight of an element in $H_i(\mathcal C)$ and $H^i(\mathcal C)$, respectively.
For trivial homology groups the associated distance is defined as $\infty$.

The logical correlations of $\mathcal F$ are recovered from the homology group, that is, for example,
\begin{equation} \label{eq:dual-log-corr}
\begin{split}
    H_{2}(\mathcal F) &\cong H_0(\mathcal R) \otimes H_{2}(\mathcal C) \oplus H_1(\mathcal R) \otimes H_{1}(\mathcal C) \\
    &= \langle
    (\mathbf 0, \mathbf 1 \otimes \ell_Z) \mid \ell_Z \in \ker H_X / \IM H_Z^T
    \rangle,
\end{split}
\end{equation}
where 
$\ell_Z$ is a logical $Z$ operator of the base code.
See~\cite{supp} for a detailed derivation.
This operator is a dual logical correlation; it acts as a copy of the logical $Z$ operator on the qubits in the $R_1 \otimes C_1$ block of the $F_1$ space.
The dual logical errors are given by 
\begin{equation} \label{eq:dual-log-error}
\begin{split}
    &H^{2}(\mathcal F) \cong H^0(\mathcal R) \otimes H^{2}(\mathcal C) \oplus H^1(\mathcal R) \otimes H^{1}(\mathcal C) \\
    &= \langle
    (\mathbf 0, (1,0,\ldots,0) \otimes \ell_X)
    \mid
    \ell_X \in \ker H_Z / \IM H_X^T
    \rangle,
\end{split}
\end{equation}
where $\ell_X$ is a logical $X$ operator of the base code.
We observe that this operator acts as a logical $X$ operator on one of the factors in the $R_1 \otimes C_1$ portion of the $F_1$ space.
Applying \cref{eq:fault_distance}, we find that the (dual) fault distance of $\mathcal F$ is $d_2 = d_X$, 
the $X$-distance of the base code.

For our choice of $\mathcal R$, the homology groups  $H_0(\mathcal R)$ and $H^0(\mathcal R)$ are trivial, so the fault complex lacks primal logical correlations arising from the base code's logical $X$.
This may seem surprising, given that the foliated surface code can be used to prepare an encoded Bell state on the two boundaries~\cite{raussendorf_long-range_2005}.
There is no contradiction, however: the interpretation of fault complexes as foliated codes implicitly assumes that all qubits are measured in the $X$ basis during the protocol, and therefore that the encoded Bell state of~\cite{raussendorf_long-range_2005} is measured destructively.
The analogous CBQC interpretation is of a memory experiment with logical state preparation and readout both performed in the $Z$ basis.
In addition, we note that all the logical correlations can be recovered by considering an alternative repetition code PCM; see~\cite{supp}.

\prlsection{Stability experiments}
Lattice surgery~\cite{horsman_surface_2012} is the leading technique for performing logical operations on topological codes~\cite{,fowler2019low,litinski_game_2019}, and stability experiments~\cite{gidney_stability_2022} estimate the logical error during a lattice surgery operation.
Such experiments test our ability to accurately measure the product of many stabilizer generators.

Returning to our foliated CSS code example, suppose that $H_X$ of the base code $\mathcal C$ is rank deficient.
The fault complex $\mathcal F = \mathcal R \times \mathcal C$ now has additional logical correlations and errors coming, respectively, from the following homology groups
\begin{equation} \label{eq:primal-log-corr-error}
\begin{split}
    H^1(\mathcal F) 
    &\cong \langle
    (\mathbf 0, (1,0,\ldots,0) \otimes g)
    \mid g \in \ker H_X^T
    \rangle,\\
    H_1(\mathcal F) 
    &\cong \langle 
    (\mathbf 0, \mathbf 1 \otimes h)
    \mid h \in C_0 / \IM H_X
    \rangle
\end{split}
\end{equation}
Here, $g$ represents any subset of $H_X$ rows that sum to zero and $h$ is any vector in $C_0$ that is not the syndrome of a $Z$-type error.
An example of such a code is a 2D surface code with four smooth boundaries; in this case
we have $g = \bf 1$ (the product of all $X$ stabilizers is $I$) and $h = (1,0,\ldots,0)$ (all valid syndromes have even weight).
The logical correlation $g$ is exactly the product of stabilizer generators considered in stability experiments.
Using \cref{eq:fault_distance} we find that the primal distance of $\mathcal F$ is $d_1 = \delta$, i.e., equal the distance of the repetition code (or the number of CBQC measurement rounds cf.~\cite{gidney_stability_2022}).
A 2D surface code with four smooth boundaries encodes no logical qubits, but if instead we let $\mathcal C$ be the 2D toric code (with periodic boundaries) then our formalism shows that the fault complex $\mathcal F = \mathcal R \times \mathcal C$ can be used to perform a combined memory and stability experiment; see~\cite{supp}.

\prlsection{Single-shot lattice surgery}
Certain CSS codes are naturally associated with length-3 or length-4 chain complexes, where the additional vector spaces represent metachecks, i.e., redundancies between subsets of checks.
Metachecks are related to single-shot error correction~\cite{bombin_single-shot_2015,campbell_theory_2019, quintavalle_single-shot_2021}, where a single round of parity-check measurements suffices for fault-tolerant QEC.

As a prototypical example of a code with metachecks, let $\mathcal C$ represent the 3D toric code~\cite{dennis_topological_2002,vasmer_three-dimensional_2019} defined on an $L\times L\times L$ cubic tiling, which has single-shot QEC for $Z$ errors. 
We focus on this code for ease of presentation, but our results also hold for true single-shot CSS codes such as the 4D toric code.
We assign the metacheck matrix $M_X = \partial_1^C$ and PCMs $H_X = \partial_2^C$ and $H_Z = (\partial_3^C)^T$.
The primal detectors of the fault complex $\mathcal F = \mathcal R \times \mathcal C$ are given by $\partial_2$, which now contains the metachecks; see~\cite{supp} for the explicit expression.
We note that $\mathcal F$ is effectively a length-3 chain complex exactly because `data' and `measurement' faults are treated on an equal footing, see~\cite{supp} for an extended discussion.

The expression for the primal logical correlations and errors of $\mathcal F$ is the same as \cref{eq:primal-log-corr-error}, but now with $g \in \ker H_X^T / \IM M_X^T$ and $h \in \ker M_X / \IM H_X$.
As in the previous example, $g$ is the logical correlation relevant for stability experiments and lattice surgery, and $h$ is the logical error that can disrupt this correlation.
In the 3D toric code, there are three independent choices of $g$ given by all edges cutting through one of the independent 2D planes of the tiling.
The corresponding $h$ vectors correspond to non-contractible chains of edges along one of the coordinate axes of the tiling.
As a result, all choices of $h$ have extensive weight, i.e., $|h| \geq L$.
Thus, even for a fault complex formed using a constant-length repetition code, a macroscopic number of faults (${\sim}L$) is necessary to disrupt the $g$ logical correlations.
This is captured by the primal fault distance of $\mathcal F$, $d_1 = \delta L$, where $\delta = d_1(\mathcal R)$ is the distance of the repetition code. 
Therefore, higher-dimensional topological codes such as the 3D and 4D toric codes are compatible with \emph{single-shot lattice surgery}, though at the cost of reduced performance.
The full distance can be restored by choosing $\mathcal R$ such that $\delta = L = O(\sqrt{d})$, (where $d$ is the distance of the code), or equivalently by performing $O(\sqrt d)$ rounds of stabilizer measurement in the lattice surgery protocol.
This should be contrasted with the 2D toric code case, where the number of measurement rounds must be $O(d)$.
Thus, we expect that a fault-tolerant quantum computing architecture based on the 4D toric code would have an asymptotic spacetime overhead reduction when compared to the standard 2D toric code architecture.

We address the case of single-shot codes with rank-deficient PCMs without metachecks in~\cite{supp}.
We note that there exist single-shot quantum codes with full-rank PCMs~\cite{fawzi_constant_2020,gu_single-shot_2024}, to which our arguments above cannot be applied, and the existence of single-shot lattice surgery protocols for these codes remains an open question.
We conjecture that codes enabling single-shot lattice surgery without performance degradation can be constructed from the balanced product of two good qLDPC codes~\cite{dinur_good_2023, leverrier_quantum_2022, panteleev_asymptotically_2022, breuckmann_balanced_2021, hastings_fiber_2021}. 

\prlsection{Improved decoding of single-shot codes}
Estimating the error threshold of a single-shot code requires simulating multiple rounds of noisy syndrome measurement until the threshold has converged~\cite{brown_fault-tolerant_2016}. 
Since in MBQC there always exist detectors spanning multiple rounds, decoding must proceed using an overlapping window decoder~\cite{dennis_topological_2002,skoric_parallel_2023,berent_analog_2024,huang_increasing_2024,bombin_modular_2023}, which is naturally defined for a fault complex.
The fault complex framework provides a systematic guide for constructing efficient simulations and evaluating the performance of these protocols.
    
\begin{figure}
    \includegraphics{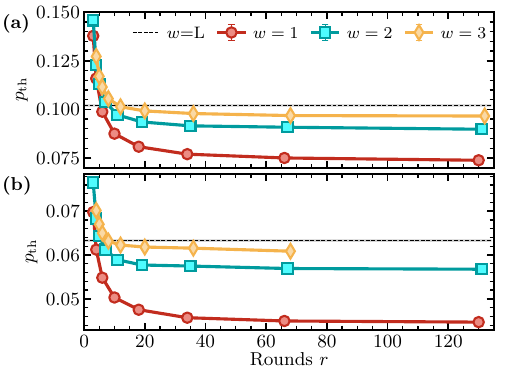}
    \caption{
    Sustainable thresholds for 3D (a) and 4D (b) toric codes under phenomenological Pauli noise. Markers represent the threshold for a $(w, 1)$-overlapping window decoder as a function of noisy syndrome rounds. The black dashed line shows the threshold for the optimal window choice $w = L$; see \cite{supp} for details. 
    }
    \label{fig:numerics}
\end{figure}

A $(w, c)-$overlapping window decoder determines in each round a correction for a window of $w \in \mathbb{N}^+$ rounds and commits a correction 
to $c \leq w$ rounds; see Refs.~\cite{huang_increasing_2024, scruby_high-threshold_2024} for a more detailed description. Here, one round constitutes a check node and a bit node layer; see Fig.~\ref{fig:surface_code_foliation}.
For the 3D toric code, the effective distance of the decoding window then becomes $\min(w L, L^2)$~\cite{supp}. However, $wL$ is the weight of time-like logical errors, and in this section, we consider memory experiments, i.e., only space-like logical errors cause logical failures; see~\cite{supp} for stability experiment simulations.
State-of-the-art results in decoding higher-dimensional topological codes~\cite{higgott_improved_2023} employ a single-stage decoding approach that is recovered using a $(1, 1)$-overlapping window decoder~\cite{supp}.
We find that increasing $w$ to $2$ or $3$ significantly increases the sustainable threshold of 3D and 4D toric codes compared to $w = 1$ when using belief propagation (BP) plus ordered statistics decoding (OSD)~\cite{panteleev_degenerate_2021,roffe_decoding_2020}; see \cite{supp} for details concerning the simulations.
For $w = 3$, the thresholds for phenomenological Pauli noise (see~\cite{supp} for a photonic noise model) of approximately $9.65\%$ (3D) and $5.9\%$ (4D) surpass all previous results~\cite{breuckmann2017local, breuckmann2018scalable, duivenvoorden_renormalization_2019, kubica2018abc, vasmer2021cellular, aloshious2021decoding, quintavalle_single-shot_2021, higgott_improved_2023, scruby_local_2023}, approaching the thresholds achieved with the optimal window choice $w = L$; see \sfigref{fig:numerics}{a} and \sfigref{fig:numerics}{b}.
We note that for the 3D toric code, we only consider primal faults, as the dual side of the fault complex does not have the single-shot property.

While larger decoding regions increase decoding time, recent advances provide solutions. Findings on localized statistics decoding~\cite{hillmann_localized_2025}, a parallelizable OSD variant, and belief propagation for higher-dimensional toric codes~\cite{scruby_local_2023} suggest time-efficient decoding with minimal performance reduction, supporting the practical applicability of our results.

\prlsection{Conclusion}
We introduced fault complexes to represent dynamic QEC protocols and characterize foliated codes.
This approach provided insights into stability experiments and lattice surgery, and yielded improved decoding protocols for higher-dimensional topological codes like the 4D toric code.
The increased thresholds and space-time overhead advantages make 4D toric codes promising for compatible architectures like photonics~\cite{bourassa_blueprint_2021, tzitrin_fault-tolerant_2021, bartolucci_fusion-based_2023, alexander2025manufacturable, walshe_linear-optical_2025, aghaee_rad_scaling_2025}, trapped ions~\cite{berthusen_experiments_2024}, and neutral atoms~\cite{bluvstein_logical_2024}.
While algorithmic fault tolerance~\cite{zhou_low-overhead_2025} offers alower overhead, it requires complex decoding; 4D toric codes allow simpler windowed decoding.
Meanwhile, the lower threshold requirements of 3D toric codes make them suitable for experimental validation in the near term.

Our formalism can extend to circuit-level noise by, for example, reformulating Li's~\cite{li_low-density_2025} LDPC representation of fault-tolerant quantum circuits (many of their results in~\cite[Sec.\ VIII]{li_low-density_2025} are already structurally equivalent to~\cref{eq:tensor_product_boundary_maps}).
Future work could explore the fault complexes of subsystem codes~\cite{kribs2005unified,kribs2006operator,poulin_stabilizer_2005} (especially those with single-shot error correction~\cite{bombin_single-shot_2015,brown_fault-tolerant_2016,kubica_single-shot_2022}), symplectic chain complexes for non-CSS codes, and alternative product constructions like balanced~\cite{breuckmann_balanced_2021} and lifted~\cite{panteleev_quantum_2022} products.
In addition, by using alternative linear codes in the construction, we may hope to algebraically recover FTGS beyond foliation~\cite{nickerson_measurement_2018,bombin_fault-tolerant_2023}.
See~\cite{supp} for some intial ideas on these extensions.

\begin{acknowledgments}
\prlsection{Acknowledgments}
The authors would like to thank Ben Brown, Christopher Chubb, Austin Daniel, Arthur Pesah, Tom Stace, Alex Townsend-Teague, and Dominic Williamson for insightful discussions.
We thank Rafael Alexander, J. Eli Bourassa, and Eric Sabo for feedback on an earlier draft of this manuscript.
Research at Perimeter Institute is supported in part by the Government of Canada through the Department of Innovation, Science and Economic Development Canada and by the Province of Ontario through the Ministry of Colleges and Universities.
Computations were performed on the Niagara supercomputer at the SciNet HPC Consortium. SciNet is funded by Innovation, Science and Economic Development Canada; the Digital Research Alliance of Canada; the Ontario Research Fund: Research Excellence; and the University of Toronto.
\end{acknowledgments}

\pagebreak

\clearpage

\onecolumngrid
\begin{center}
\textbf{\Large Supplementary Information for \\ ``Single-shot and measurement-based quantum error correction via fault complexes"}
\end{center}
\vspace{1em}
\twocolumngrid

\setcounter{equation}{0}
\setcounter{figure}{0}
\setcounter{table}{0}   
\setcounter{page}{1}
\setcounter{section}{0}
\makeatletter

\renewcommand{\thesection}{S\arabic{section}}
\renewcommand{\theequation}{S\arabic{equation}}
\renewcommand{\thefigure}{S\arabic{figure}}
\renewcommand{\thetable}{S\arabic{table}}

\section{Derivation of logical correlations and errors \label{sec:derivations}}

Here, we derive the expressions for logical errors/correlations and fault distances given in the main text and discuss their interpretation in more detail.
We consider the fault complex $\mathcal F = \mathcal R \times \mathcal C$, where $\mathcal C$ describes a CSS code and $\mathcal R$ describes the repetition code whose parity-check matrix has a redundant row\footnote{One can think of this parity-check matrix as being the vertex-to-edge adjacency matrix of the cycle graph.}, as the main text considers a special case of this. Explicitly, 
\begin{equation}
    R = \begin{bmatrix}
        1 & 1 & 0 & \ldots & 0 & 0 \\
        0 & 1 & 1 & \ldots & 0 & 0 \\
        \vdots & & & \ddots & & \vdots\\
        0 & 0 & 0 & \ldots & 1 & 1 \\
        1 & 0 & 0 & \ldots & 0 & 1 \\
    \end{bmatrix}.
\end{equation}
We will refer to the $i^\text{th}$ boundary operator of $\mathcal C$ by $\partial_i$ to ease notational clutter.

To obtain the homology groups of the fault complex, we make use of the K\"{u}nneth formula:
\begin{equation} \label{eq:Kunneth-hom}
    H_j(\mathcal F) \cong H_0(\mathcal R) \otimes H_j(\mathcal C) \oplus H_1(\mathcal R) \otimes H_{j-1}(\mathcal C).
\end{equation}
Similarly, for the cohomology groups, we have
\begin{equation} \label{eq:Kunneth-cohom}
    H^j(\mathcal F) \cong H^0(\mathcal R) \otimes H^j(\mathcal C) \oplus H^1(\mathcal R) \otimes H^{j-1}(\mathcal C).
\end{equation}
For $\mathcal R$, one can verify that
\begin{equation} \label{eq:homology-R}
\begin{split}
    H_0(\mathcal R) &= R_0 / \IM R, \\ 
    H_1(\mathcal R) &= \ker R / \{ \mathbf{0} \} \cong \ker R, \\
    H^0(\mathcal R) &= \ker R^T / \{ \mathbf 0 \} \cong \ker R^T, \\
    H^1(\mathcal R) &= R_1 / \IM R^T,
\end{split}
\end{equation}
Further calculation shows that
\begin{equation}
\begin{split}
&\IM R = \IM R^T \\ &= \SPAN \{ (1,1,0,\ldots,0), \ldots, (0,\ldots,0,1,1) \}
\end{split}
\end{equation}
and $\ker R = \ker R^T = \{ \mathbf 0, \mathbf 1 \}$.
Therefore
\begin{equation} \label{eq:hom-rep-redundant}
\begin{split}
    H_0(\mathcal R) &\cong \{ \mathbf 0, (1,0,\ldots,0) \}, \\ 
    H_1(\mathcal R) &\cong \{ \mathbf 0, \mathbf 1 \}, \\
    H^0(\mathcal R) &\cong \{ \mathbf 0, \mathbf 1 \}, \\
    H^1(\mathcal R) &\cong \{ \mathbf 0, (1,0,\ldots,0) \}.
\end{split}
\end{equation}

Using \cref{eq:Kunneth-hom,eq:hom-rep-redundant} we have 
\begin{equation} \label{eq:primal-log-error}
\begin{split}
    H_i(\mathcal F) &\cong H_0(\mathcal R) \otimes H_i(\mathcal C) \oplus H_1(\mathcal R) \otimes H_{i-1}(\mathcal C) \\
    &= \langle ((1,0,\ldots,0) \otimes \ell_Z , \mathbf 0),
    (\mathbf 0, \mathbf 1 \otimes h)
    \rangle, 
\end{split}
\end{equation}
where $\ell_Z \in \ker \partial_i / \IM \partial_{i+1}$ and $h \in \ker \partial_{i-1} / \IM \partial_i$.
Similarly, we have
\begin{equation} \label{eq:primal-log-corr}
\begin{split}
    H^i(\mathcal F) &\cong H^0(\mathcal R) \otimes H^i(\mathcal C) \oplus H^1(\mathcal R) \otimes H^{i-1}(\mathcal C) \\
    &= \langle (\mathbf 1 \otimes \ell_X, \mathbf 0),
    (\mathbf 0, (1,0,\ldots,0) \otimes g)
    \rangle,
\end{split}
\end{equation}
where $\ell_X \in \ker \partial_{i+1}^T / \Im \partial_i^T$ and $g \in \ker \partial^T_i / \IM \partial_{i-1}^T$.
We refer to elements of $H^i(\mathcal C)$ as primal logical correlations and to elements of $H_i(\mathcal C)$ as primal logical errors.
Analogously for the other relevant homology groups, we have
\begin{equation} \label{app:dual-log-corr}
\begin{split}
    H_{i+1}(\mathcal F) &\cong H_0(\mathcal R) \otimes H_{i+1}(\mathcal C) \oplus H_1(\mathcal R) \otimes H_{i}(\mathcal C) \\
    &= \langle ((1,0,\ldots,0) \otimes g' , \mathbf 0),
    (\mathbf 0, \mathbf 1 \otimes \ell_Z')
    \rangle, 
\end{split}
\end{equation}
where $g' \in \ker \partial_{i+1} / \IM \partial_{i+2}$ and $\ell_Z' \in \ker \partial_i / \IM \partial_{i+1}$.
And
\begin{equation} \label{app:dual-log-error}
\begin{split}
    H^{i+1}(\mathcal F) &\cong H^0(\mathcal R) \otimes H^{i+1}(\mathcal C) \oplus H^1(\mathcal R) \otimes H^{i}(\mathcal C) \\
    &= \langle (\mathbf 1 \otimes h', \mathbf 0),
    (\mathbf 0, (1,0,\ldots,0) \otimes \ell_X')
    \rangle,
\end{split}
\end{equation}
where $h' \in \ker \partial_{i+2}^T / \IM \partial_{i+1}^T$ and $\ell_X' \in \ker \partial^T_{i+1} / \IM \partial_i^T$.
We refer to the elements of $H_{i+1}(\mathcal F)$ as dual logical correlations and to $H^{i+1}(\mathcal F)$ as dual logical errors.

Recall that the number of primal and dual logical correlations (or errors) of a fault complex are defined to be $k_i = \dim H_i(\mathcal F)$ and $k_{i+1} = H_{i+1}(\mathcal F)$, respectively.
Using Eq.~\eqref{eq:Kunneth-hom}, we obtain the general expression for these quantities
\begin{equation*}
\begin{split}
    k_i&= \dim H_0(\mathcal R) \dim H_{i}(\mathcal C) + \dim H_1(\mathcal R) \dim H_{i-1}(\mathcal C),\\
    k_{i+1} &= \dim H_0(\mathcal R) \dim H_{i+1}(\mathcal C) + \dim H_1(\mathcal R) \dim H_{i}(\mathcal C),
\end{split}
\end{equation*}
which in our case further simplifies via Eq.~\eqref{eq:hom-rep-redundant}, giving
\begin{equation}
\begin{split}
    k_i&= \dim H_{i}(\mathcal C) + \dim H_{i-1}(\mathcal C),\\
    k_{i+1} &= \dim H_{i+1}(\mathcal C) + \dim H_{i}(\mathcal C),
\end{split}
\end{equation}
We see that $k_i \neq k_{i+1}$ in general, with equality only when $\dim H_{i-1} = \dim H_{i+1}$.

Let us now consider the interpretation of the logical correlations and errors in terms of the CSS code described by $\mathcal C$.
We concentrate on the primal case as the dual case is analogous.
Note that each $\ell_Z$ describes the support of a logical $Z$ operator (of the CSS code) and each $\ell_X$ describes the support of a logical $X$ operator.
Therefore, symplectic pairs of logical operators in the CSS code lift to symplectic pairs of logical correlations and errors in the fault complex.
If $\partial_i$ is full rank, then $\IM \partial_i = C_{i-1}$ and $\ker \partial_i^T = \{ \mathbf 0 \}$.
In this case, the only possible choices for $g$ and $h$ are the zero vector, and therefore the generators of the homology groups all derive from the logical operators of the CSS code.
Now suppose that $C_i$ is rank-deficient.
In this case, $g$ describes a subset of $X$-type stabilizer generators whose product is the identity, but which is not an $X$-type meta-check of the code.
And $h$ is a vector in $C_{i-1}$ with trivial meta-syndrome, but which is not the syndrome of a $Z$-type error
(see the main text for an example).
If the code does not have $X$-type meta-checks (i.e., $\mathcal C$ is length-2) then $g$ can be any subset of $X$-type stabilizer generators with trivial product, and $h$ can be any vector in $C_{i-1}$ that is not the syndrome of a $Z$-type error.

To give an explicit example, suppose that $\mathcal C$ describes the 2D toric code with periodic boundaries and the usual stabilizer generators ($X$-type star operators and $Z$-type plaquette operators).
This code has two encoded qubits and two non-trivial stabilizer relations as the product of all the stabilizer generators of the same type is the identity.
In addition, every error syndrome must have even weight (this is sometimes called a materialized symmetry of the code~\cite{Brown2022Generalized}). 
Therefore, we can take $g = \mathbf 1$ and $h = (1,0,\ldots,0)$.

We note that one can always construct a metacheck matrix for a rank-deficient parity-check matrix by Gaussian elimination.
However, at least in topological codes, the metachecks are usually defined to be the relations comprising $O(1)$ stabilizer generators, i.e., the metacheck matrix has low-weight rows (and low-weight columns as it turns out).
On the other hand, the spacelike invariants such as $g$ and $g'$ above are relations comprising an extensive number of stabilizer generators.
The rationale for this distinction is that if each stabilizer measurement is noisy, then combining an extensive number of measurement outcomes will yield no useful information in the limit of large code size.
Indeed, this is what makes it non-trivial to preserve the value of these logical correlations in QEC protocols.

Let us also comment explicitly on the case of a single-shot CSS code with rank-deficient parity-check matrices that is completely described by a length-2 chain complex. 
Here, there is no explicit notion of metachecks.
As noted above, metachecks can be constructed based on the linear dependencies in the parity-check matrices.
The resulting metacheck matrices can be interpreted as classical codes with distances $d_{M_X}$ and $d_{M_Z}$.
Our results therefore extend to codes where $d_{M_X}$ and $d_{M_Z}$ are non-trivial.
However, such procedures generally lead to higher-weight rows.
Determining meta-checks of minimal weight is expected to be (at least) NP-hard, in analogy to determining the minimum distance of a classical code.

\subsection{Fault distances}

The fault distances of $\mathcal F$ can essentially be read off~\cref{eq:primal-log-error,app:dual-log-error}, but we provide a derivation here for completeness.
For an arbitrary length-1 chain complex $\mathcal R$, Zeng and Pryadko showed in Ref.~\cite{zeng_higher-dimensional_2019} that the product complex $\mathcal F = \mathcal R \times \mathcal C$ has homological distances
\begin{equation}
\label{eq:hom_distance}
\begin{split}
    d_{j}(\mathcal F) &= \min[d_0(\mathcal{R})d_i(\mathcal{C}), d_1(\mathcal{R})d_{i-1}(\mathcal{C})],  \\
    d_{j}(\mathcal{F}^T) &= \min[d_0(\mathcal{R}^T)d_{j}(\mathcal{C}^T), d_1(\mathcal{R}^T)d_{j-1}(\mathcal{C}^T)],
\end{split}
\end{equation}
where $d_j(\mathcal C)$ ($d_j(\mathcal F^T)$) is equal to the weight of the minimal weight operator in the (co)homology group $H_j(\mathcal C)$ ($H^j(\mathcal C)$).
For $\mathcal R$ above we have
\begin{equation}
\begin{split}
    d_0(\mathcal R) = d_1(\mathcal R^T) = 1,\\
    d_1(\mathcal R) = d_0(\mathcal R^T) = \delta,
\end{split}
\end{equation}
where $\delta$ is the length of the repetition code.
Therefore the primal and dual distances of $\mathcal F$ are
\begin{equation}
\label{app:fault_distance}
\begin{split}
    d_{i} &= \min[d_i(\mathcal{C}), \delta d_{i-1}(\mathcal{C})],  \\
    d_{i+1} &= \min[\delta d_{i+1}(\mathcal{C^T}), d_{i}(\mathcal{C^T})].
\end{split}
\end{equation}

\section{Single-shot decoding for fault complexes}

Here we give further details on the decoding of fault complexes with the single-shot property and explain a key difference between the fault complex and CBQC pictures of single-shot decoding.

We consider the fault complex $\mathcal F = \mathcal R \times \mathcal C$, where $\mathcal R$ describes a distance $\ell$ repetition code with the parity-check matrix $R$ and $\mathcal C$ is a length-4 chain complex describing a CSS code with meta-checks.
For a CSS code of this type, we have
\begin{equation}
    \begin{split}
        H_X &= \partial_2^C, \quad H_Z^T = \partial_3^C,\\
        M_X &= \partial_1^C, \quad M_Z^T = \partial_4^C,
    \end{split}
\end{equation}
where $M_X$ and $M_Z$ denote the meta-check matrices for $X$ and $Z$ checks, respectively.

Recall that the fault complex is formally a length-5 chain complex
\begin{align}
     \mathcal{F} = F_5 \stackrel{\partial_5}{\rightarrow}  F_4 \stackrel{\partial_4}{\rightarrow} F_3 \stackrel{\partial_3}{\rightarrow} F_2 \stackrel{\partial_2}{\rightarrow} F_1 \stackrel{\partial_1}{\rightarrow} F_0,
\end{align}
with boundary maps given by 
\begin{align}
\label{app:tensor_product_boundary_maps}
    \partial_{j} =  \left( \begin{array}{c|c}
         \mathbbm{1}_{r}  \otimes \partial^C_{j}  & R \otimes \mathbbm{1}_{n_{j-1}}  \\ \hline
         0 & \mathbbm{1}_{c}  \otimes  \partial^C_{j-1} 
    \end{array} \right).
\end{align}
We identify the primal and dual qubits with the vector spaces $C_2$ and $C_3$, respectively, 
Hence, the graph state connectivity is specified by $\partial_3$, while the primal and dual detectors are given by $D_X = \partial_2$ and $D_Z = \partial_4^T$, respectively.

Per the main text, this fault complex has no ``meta-detectors'' and can be reduced to a length-3 sub-complex without any loss of information.
To understand why this is the case, first notice that the detector matrices $D_X$ and $D_Z$ already contain fault locations, using the terminology of CBQC, of data and ancilla faults.
This can be seen by constructing, e.g., the matrix $D_X$ for $\delta=2$ and performing column and row operations
\begin{align}
    D_X &= \left( 
    \begin{array}{ccccc}
        {\color{teal} M_X} & & & &  \\
         {\color{teal} \mathbbm{1}} & {\color{teal}H_X} & {\color{teal}\mathbbm{1}} \\
         & & {\color{teal}M_X} & &  \\
         & &  \mathbbm{1} & {\color{Maroon} H_X} & {\color{Maroon} \mathbbm{1}} \\
         & & & & {\color{Maroon} M_X}
    \end{array} \label{eq:foliated_single_shot_check}
    \right).
\end{align}
In the above matrix, the odd rows detect ancilla faults and the even rows detect data faults, while odd columns are ancilla fault locations and even columns are data fault locations.
Unlike in the CBQC framework for single-shot error correction, in the fault complex picture, there is no distinction between data faults and ancilla faults, or equivalently between detectors arising from stabilizers ($H_X$) and detectors arising from meta-checks ($M_X$).
Thus, for any error $e_X$ with syndrome $s_X = D_X e_X$ we have $\partial_1 s_X = \partial_1 \partial_2 e_X = 0$.
This distinction runs deep in the CBQC community and has historically influenced decoder design, particularly in phenomenological noise simulations where faults are sampled in separate steps.
In such simulations, faults on data qubits are first drawn from the space 
$C_i$ , and only afterwards are ancilla qubit faults sampled from $C_{i \pm 1}$.
This sequential treatment reinforces the mindset that data and ancilla faults are fundamentally distinct and should be handled separately.
Consequently, decoder designs—especially for single-shot codes with metachecks—have traditionally relied on heuristics to merge these distinct fault types into an effective decoding matrix-\cite{quintavalle_single-shot_2021, higgott_improved_2023}.
By contrast, the fault complex framework eliminates this conceptual divide.
By embedding data and measurement faults in the same vector spaces, $F_i$ and $F_{i+1}$ for primal and dual faults, respectively, it provides a natural and systematic way to construct the appropriate detector check matrices.
This is illustrated in\cref{eq:foliated_single_shot_check} where the structure of $D_X$ naturally recovers the effective parity-check matrix for decoding.
Rather than relying on intuition, as was done in prior works, the fault complex formalism directly specifies the correct boundary operators, yielding the appropriate check structure from first principles.
The first block (teal) is the relevant check matrix for a single round of the overlapping window decoder, specifying a single detector block in the overlapping window decoder for $w=1$.
The second block (red) is identical to the single-stage decoding matrix proposed in Ref.~\cite{higgott_improved_2023}. 
This block can be recovered from block one by treating the ancilla faults from the previous round (first column of block one) as perfect, which eliminates them from the decoding problem.
The full block is the relevant detector check matrix for a window size $w = 2$.

Thus, the fault complex formalism provides a rigorous derivation of the single-shot decoding strategy—resolving prior uncertainties about how to construct the appropriate parity-check matrix for decoding single-shot codes with metachecks.

\subsection{Insights on decoding}
Formalizing the simulation framework in the language of fault complexes yielded important insights in the improved decoding of single-shot code with metachecks.
In the following, we will explain based on the example of 4D toric code how those insights were obtained from analyzing the fault complex of the code.

The 4D toric code of linear lattice size $L$ has code distance $d_C = L^2$ and metacheck distance $d_M = L$.
A code with distance $d$ can correct up errors up to weight $\lfloor (d-1) / 2 \rfloor$.
Then, over a window size $w \in \mathbb{N}$ the distance of the corresponding fault complex $d(\mathcal{F})$ according to Eq.~\eqref{app:fault_distance} is given by $d(\mathcal{F}) = \min[wL, L^2]$.
Then, for $w=1$ and $L$ odd, one would not expect $L+1$ to have improved error suppression over $L$ as the weight of correctable errors is identical.
However, here and in Ref.~\cite{higgott_improved_2023} numerical results have shown that $L + 1$ does indeed improve performance, which appears paradoxical from this perspective.

Our formalism allowed us to understand this apparent paradox by recognizing that the distance $\min[wL, L^2]$ has contributions from logical correlations of different roles.
First, the emph{time-like} distance, which scales as $wL$, and the \emph{space-like} distance, which scales as $L^2$.

In Ref.~\cite{higgott_improved_2023} as well as in Figure 2 of the main text, memory experiments are performed which only space-like logical errors cause logical failures.
Nevertheless, also in memory experiments, time-like errors can occur and accumulate over repeated rounds of error correction but remain unproblematic as long as they are not converted into space-like errors. 
However, the decoder can convert accumulated time-like errors into space-like errors.
This conversion process becomes significant when analyzing single-shot thresholds, necessitating simulations over many rounds of error correction to observe threshold convergence.

We find that convergence of threshold estimates is much slower for $w = 1$, where the time-like distance is minimal. 
Due to the increased time-like distance for $w > 1$, when a time-like fault is converted into a space-like fault, this fault is more likely to directly cause a logical failure.
This is the intuitive explanation for the observation of the  faster convergence of the threshold.
In the case $w = L$, the time-like and space-like distances become identical, and convergence is immediate and it is not necessary to simulate multiple error correction rounds in principle.
However, due to boundary effects, even in that case convergence should be checked carefully.

\section{Numerical simulations}
Here we describe in more detail how the simulations are performed to obtain the sustainable thresholds of the 3D and 4D toric codes. All simulations in the manuscript were performed using a private version of \textsf{FlamingPy}~\cite{FlamingPy}.

\subsection{Phenomenological noise}
We begin by describing the case of phenomenological noise. 
Since we are only considering a single type of Pauli error in our simulations, we only need to decode a single side of the fault complex.
First, we construct the associated fault complex through foliation based on the base code, e.g., the 3D toric code, over $\delta$ rounds.
In the cluster state picture, this corresponds to a cluster state of $2\delta + 1$ layers.
As we are using 30 iterations of the min-sum belief propagation (BP) algorithm, potentially followed by ordered statistics decoding post-processing with an order-60 combination sweep (CS-60) search strategy, we do not require the gauge group generators.
Given the dual detectors $D_Z$, we sample faults corresponding to each matrix column from an i.i.d noise model with probability $p$.
Any faults occurring on the initial and final layer are removed to assert that we are starting and ending in a code state.
This assumption is equivalent to the usual assumption in the phenomenological noise model simulations that the final syndrome measurement is noise-free.
For each trial, the sampled fault vector is sliced up according to the $(w, 1)$-overlapping window decoder before being passed to the local decoder.
A more explicit description of this can be found in Refs~\cite{huang_increasing_2024, scruby_high-threshold_2024}.
After the full correction $c_{\delta}$ over $\delta$ rounds is calculated, we check whether the residual error is a logical error.
For the 3D and 4D toric codes that possess 3 and 6 logical qubits, we count as a failure if at least a single logical qubit remains in error.

\subsection{Photonic GKP noise}
\begin{figure}[!b]
    \centering
    \includegraphics{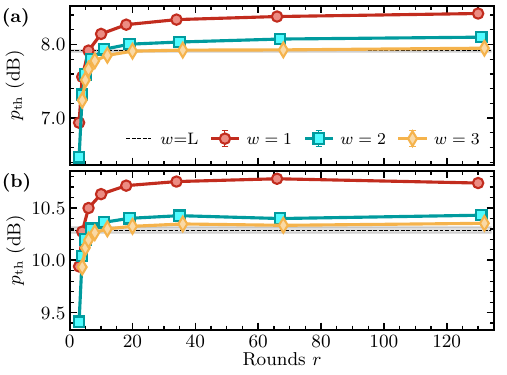}
    \caption{Sustainable thresholds for 3D (a) and 4D (b) toric codes under a noise model for a photonic GKP-based architecture. Markers represent the threshold for a $(w, 1)$-overlapping window decoder as a function of noisy syndrome rounds. The black dashed line shows the threshold for the optimal window choice $w = L$. In the 3D case, we only consider primal faults, whereas in the 4D case, we consider both primal and dual faults.}
    \label{fig:gkp_numerics}
\end{figure}

Additionally, we investigated the threshold under a noise model for Xanadu’s photonic architecture based on Gottesman-Kitaev-Preskill (GKP) qubits~\cite{gottesman_encoding_2001,
bourassa_blueprint_2021, tzitrin_fault-tolerant_2021}, with cluster state construction and error model details provided below and in~\cite{walshe_linear-optical_2025, aghaee_rad_scaling_2025}.
For the 4D toric code, we observe a threshold of approximately $10.35$ dB, comparable to that of the 2D variant~\cite{bourassa_blueprint_2021, tzitrin_fault-tolerant_2021}. 
The 3D code achieves a threshold of approximately $7.95$ dB.
In both cases, the $w = 3$ decoder archives thresholds comparable to the $w = L$ decoder, see \sfigref{fig:gkp_numerics}{a} and \sfigref{fig:gkp_numerics}{b}. 

\begin{figure*}[ht]
    \centering
    \includegraphics{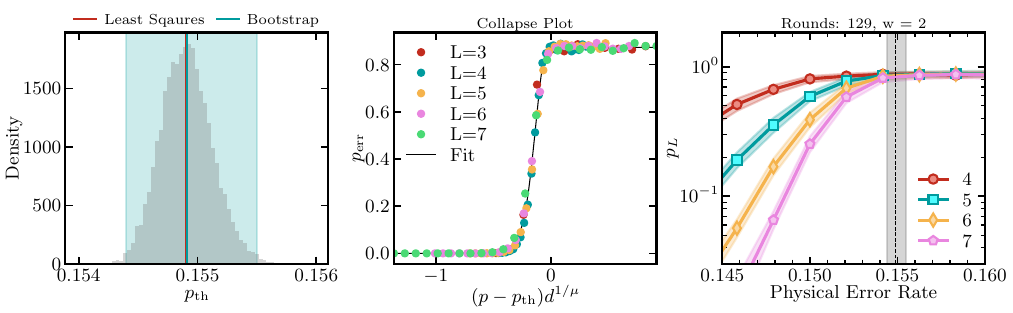}
    \caption{Example of the threshold fitting procedure for the 3D toric code under the photonic GKP noise model with an $w=2$ overlapping window decoder after $129$ rounds of noisy syndrome measurements. \textbf{Left:} Histogram of the bootstrap resamples, with the mean and $99\%$ confidence interval highlighted in blue.
    \textbf{Center:} Collapse plot of the fitted data.
    \textbf{Right:} Threshold plot of the data obtained from the sustainable threshold simulation with the threshold value highlighted as a black dashed line and $99\%$ confidence intervals as a grey shading.
    }
    \label{fig:fitting_procedure}
\end{figure*}

When simulating Xanadu's photonic architecture based on optical Gottesman-Kitaev-Preskill qubits~\cite{bourassa_blueprint_2021, tzitrin_fault-tolerant_2021, walshe_linear-optical_2025, aghaee_rad_scaling_2025}, a few adjustments to the simulation routine are necessary.
First, as we are aiming to simulate an approximation of the physical noise model, we now require knowledge about the bi-adjacency matrix of the (qubit-level) fault-tolerant graph state.
As described in Ref.~\cite{walshe_linear-optical_2025, aghaee_rad_scaling_2025}, using the process of \emph{macronization} and \emph{stiching} the qubit-level graph state can equivalently be described by several GKP Bell pairs and linear-optical circuit elements, e.g., phase shifters and beam splitters.
For each optical mode describing a GKP state, we sample a random displacement from the normal distribution $\mathcal{N}(0, \sigma^2)$ with variance $\sigma^2$ and propagate the sampled displacement through the linear optical circuit until the final homodyne measurement stage.
Computationally, this propagation of sampled displacements amounts to a series of matrix vector multiplications that introduce local correlations between the modes.
This is an effective model to describe a finite-energy GKP qunaught state with $-10 \log_{10}(\sigma^2/\sigma_\text{vac}^2)\; \mathrm{dB}$ of per-peak squeezing, where we set the vaccum variance $\sigma_\text{vac}^2 = 2$.
As described in more detail in Refs.~\cite{bourassa_blueprint_2021, tzitrin_fault-tolerant_2021, walshe_linear-optical_2025, aghaee_rad_scaling_2025}, the sampled displacements can be processed to yield measurement outcomes for the individual modes when they are measured in the \emph{reduction} step described in Ref.~\cite{walshe_linear-optical_2025}.
To improve the logical error suppression and the observed threshold, we employ as an inner decoder on the level of the GKP code an algorithm different from the binning strategy mentioned in Refs.~\cite{bourassa_blueprint_2021, tzitrin_fault-tolerant_2021}.
Our correlation-aware inner decoder, described in detail in Ref.~\cite[see the supplementary information Section VI.A]{aghaee_rad_scaling_2025}, reduces the effective error probability of individual qubits in the cluster state.
From this, an effective qubit-level noise model can be obtained and the rest of the simulation routine continues as described in the previous section.

\subsection{Threshold estimation}
The common way of obtaining an estimate of the threshold based on the simulation of finite system sizes consists of performing a quadratic fit with the rescaled error rate~\cite{wang_confinement-higgs_2003, harrington_analysis_2004}, that is,
\begin{align}
    x &= (p - p_{th}) d^{1 / \mu},  \label{eq:rescaled_physical_error_rate}\\
    f(x)  &= a x^2 + b x + c,
\end{align}
with  $p_{th}$ the threshold to be fitted, $d$ the code distance, $\mu$ a scaling parameter and $a, b, c$ free parameters of the quadratic polynomial.
However, this approach cannot be followed for determining the threshold estimate of sustainable threshold simulations. For a sufficiently large number of decoding rounds, the observed logical error rate will saturate at $p_L = 1 - (0.5)^k$ in the vicinity of the threshold, i.e., for $p \approx p_{th}$.
Thus, it is impossible to observe a threshold as a crossing of curves for finite system sizes: this crossing lies in a region where the error rates saturate.
To capture accurately the behavior described above, we consider fitting to an alternative model, that is,
\begin{equation} \label{eq:fit_model}
    g(x) = a \left[1 - \left(1 - \frac{1 + \tanh( b x )}{2} \right)^c\right],
\end{equation}
where $x$ is the rescaled error rate as in Eq.~\eqref{eq:rescaled_physical_error_rate} and $a, b, c$ are free fit parameters.
In principle, if the fit is performed over a sufficient range of $x$, one does not need to fit $a$ and can set it instead to its analytical value $a \equiv 1 - (0.5)^k$; however, we leave $a$ as a free parameter. The parameter $c$ typically scales with the number of decoding rounds.

To estimate the threshold based on the model Eq.~\eqref{eq:fit_model}, we use the bootstrap resampling technique with $10000$ resamples; see~\cite{threshold_bootstrap} for an implementation.
The obtained (asymmetric) error bars represent the $99\%$ confidence intervals.
\figref{fig:fitting_procedure} contains some example plots for the threshold fitting procedure.

\subsection{Stability experiments}
\begin{figure}[!t]
    \centering
    \includegraphics{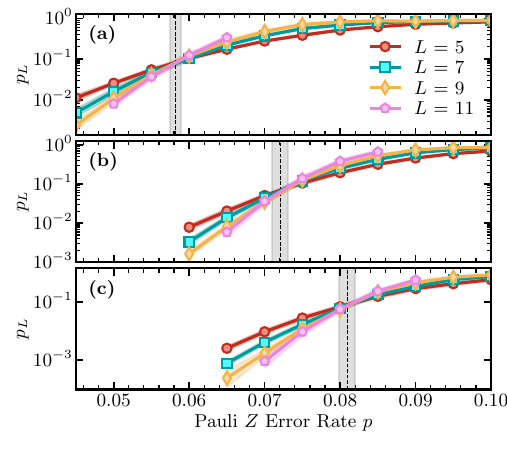}
    \caption{Logical error rate for the time-like logical correlation of a 3D toric code for different window sizes of the overlapping window decoder. In \textbf{(a)} $w = 1$,
    \textbf{(b)} $w = 2$,
    and \textbf{(c)} $ w = 3$.
    Vertical black dashed lines indicate the fitted threshold.
    The shading indicates a confidence interval, indicating hypotheses whose likelihoods are within a factor of 1000 of the maximum likelihood estimate.
    }
    \label{fig:stability_experiment}
\end{figure}

\begin{figure}[ht]
    \centering
    \includegraphics{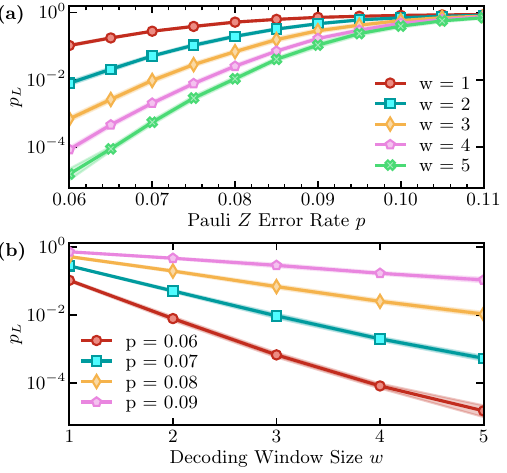}
    \caption{Stability experiments with the $L=5$ 3D toric code.
    \textbf{(a)} Logical error rate as a function of the Pauli error rate $p$ for various sizes of the decoding window $w$.
    \textbf{(b)} Scaling of the logical error rate below the threshold as a function of the size of the decoding window $w$.
    }
    \label{fig:stability-window}
\end{figure}

To further support our claim that 3D and 4D toric codes are compatible with single-shot lattice surgery, we perform simulations of stability experiments.
Stability experiments can be used as a proxy for simulating lattice surgery because they probe the accuracy of reliably inferring the outcomes of an extensive number of stabilizers~\cite{gidney_stability_2022}.
We perform numerical simulations over $32$ decoding rounds for a 3D toric code of linear size $L$ (with periodic boundaries in time to avoid finite-size effects).
In this case, the corresponding fault complex is equivalent to the chain complex that describes an asymmetric 4D toric code with dimensions $L\times L \times L\times w$ for each decoding window.
Decoding proceeds with an $(w, 1)$ overlapping window decoder.
In this case, the effective primal fault distance is $d_1 = w L$, i.e., for $w \leq L$ the distance is limited by the size of the decoding window.
But even for $w = 1$ we expect to observe an error threshold based on the distance scaling $d \sim L$.
The $w = 1$ case corresponds to single-shot lattice surgery, but we note that any $w = O(1)$ corresponds to constant-time overhead lattice surgery.
Our results are summarized in \figref{fig:stability_experiment}.
As expected from the distance calculation, we observe an error threshold for all values of $w$, with the logical error rate decreasing with increasing $L$ below the threshold.
By increasing the decoding window size $w$ from $w = 1$ in \sfigref{fig:stability_experiment}{a}, to larger, but constant, values of $w = 2$ in \sfigref{fig:stability_experiment}{b} and $w = 3$ in \sfigref{fig:stability_experiment}{c} we observe increased error thresholds in the stability experiments.
We note that choosing $w$ adaptively on the code size, e.g., $w = L$, balances time-like and space-like errors.
In practice, this allows one to find a trade-off between the number of syndrome measurement rounds required for lattice surgery, and thus the time overhead for logical gates, and the achievable logical error rates of the gate.
We emphasize that these observations are in stark contrast to stability experiments with non-single-shot codes, as performed by Gidney~\cite{gidney_stability_2022} on the 2D surface code, where the effective distance is independent of $L$.
For two codes with $L_2 > L_1$, the code with $L_1$ will always have a lower error rate than the code with $L_2$ if decoded over the same window size $w$.
Additionally, as expected, we observe exponential suppression of the logical error rate below threshold; see \figref{fig:stability-window}.

In the main text, we also proposed a joint memory and stability experiment for the 2D toric code with periodic boundary conditions (this would also work in higher dimensions).
This experiment proceeds in the same way as a memory experiment, namely
\begin{enumerate}
    \item Initialize all qubits in the $|0\rangle$ state.
    \item Measure the stabilizers $T$ times.
    \item Measure all qubits in the $Z$ basis.
\end{enumerate}
After decoding, we can then estimate the final values of the logical $Z$ operators, and the product of all the $X$ stabilizers (in a single round).
These observables correspond, respectively, to the logical correlations of the fault complex descending from the logical $Z$ operators of the 2D toric code (memory), and to the logical correlation descending from the redundancy in $H_X$ (stability).
Each of these observables should be deterministic in the absence of errors, so using this experiment, we can estimate the probability of a spacelike logical error that disrupts the memory observable or a timelike logical error that disrupts the stability observable.
We can vary the code distance $d$ and the number of rounds $T$ in order to estimate the threshold for both types of observable.

\section{Fault complexes of subsystem codes}

Here, we briefly discuss the extension of our fault complex formalism to subsystem codes~\cite{kribs2005unified,kribs2006operator,Poulin2005Subsystem}.
We recall that a CSS subsystem code is defined by two (gauge) matrices $G_X$ and $G_Z$, but unlike in the subspace code case, we do not require that $G_X G_Z^T = 0$~\cite{Liu2024subsystemcsscodes}.
The $X$-type stabilizers of the subsystem code are the elements of the row space of $G_X$ that have even overlap all rows of $G_Z$, and vice versa for the $Z$-type stabilizers.
We use the notation $H_X$ and $H_Z$ for the $X$- and $Z$-type stabilizer matrices, respectively.

Prior work on foliation~\cite{bolt_foliated_2016,bolt_decoding_2018,brown_universal_2020} gives us clues for the correct form of the fault complexes of subsystem codes.
We know that in a foliated CSS subsystem code, the ancilla qubit connections correspond to the rows of $G_X$ and $G_Z$, and the data qubits connections are unchanged.
This suggests that---for a length-3 fault complex---the boundary operator $\partial_2$ describing the cluster state connectivity should be as in \cref{app:tensor_product_boundary_maps} with $\partial_i^C = G_Z^T$ and $\partial_{i-1}^C = G_X$.
However, for the other boundary operators, we should use $H_X$ and $H_Z$ instead of $G_X$ and $G_Z$, as subsystem code detectors are formed from the stabilizers, not the checks.
Explicitly,
\begin{align}
    \label{eq:fault-complex-primal-det-mat}
    D_X = \partial_{1} &=  \left( \begin{array}{c|c}
         \mathbbm{1}_{r}  \otimes H_X  & R \otimes \mathbbm{1}_{n_{0}} 
    \end{array} \right), \\
    \partial_{2} &= \left( \begin{array}{c|c}
        \mathbbm{1}_{r} \otimes G_Z^T & R \otimes \mathbbm{1}_{n_{1}} \\ \hline
        0 & \mathbbm{1}_{c} \otimes G_X
    \end{array} \right), \\
    D_Z^T = \partial_{3} &= \left( \begin{array}{c}
        R \otimes \mathbbm{1}_{n_{2}} \\ \hline
        \mathbbm{1}_{c} \otimes H_Z^T
    \end{array} \right).
\end{align}
For the case of CSS subsystem codes with meta-checks, we expect that $D_X$ and $D_Z$ would simply need to include $M_X$ and $M_Z$ (the meta-check matrices) and would resemble \cref{eq:foliated_single_shot_check}.
Given the similarity of the detector matrices, the decoding problem for these codes should be identical in structure to the decoding problem for fault complexes we have previously covered.
It would be, therefore, interesting to examine the performance of, e.g., subsystem codes with single-shot error correction~\cite{bombin_single-shot_2015,brown_fault-tolerant_2016,kubica_single-shot_2022,stahl_single-shot_2024} using the approach explored in this work.

The changes outlined above are relatively minor, and we therefore anticipate that foliated CSS subsystem codes can be represented naturally as fault complexes.
The main barrier to formalizing our intuition is finding the appropriate representation of CSS subsystem codes as chain complexes (see~\cite{zeng_minimal_2020,kubica_single-shot_2022} for some possibilities).
We leave this for future work.

\section{Fault complexes of circuit models}
In Ref.~\cite{li_low-density_2025}, Li proposes a low-density parity-check representation of fault-tolerant quantum circuits. 
The formalism introduced there can be seen as another perspective of space-time codes; see also Refs.~\cite{bacon_sparse_2017, delfosse_spacetime_2023, gottesman_opportunities_2022, fu_error_2024}.
In Sec. VIII of~\cite{li_low-density_2025}, the author gives explicit expressions obtained from their formalism for operations that do not couple $X$ and $Z$ type operators, resulting in two disjoint Tanner graphs.
In particular, they obtain the matrix representation, for what we call the gauge group generators, as
\begin{align}
    G_X &= \left( \begin{array}{cc}
          \vec{a}_X \otimes \mathbbm{1}_n & 1_{m_X^C} \otimes \mathbf{H}_X^T  \\
         \vec{d}_X^T \otimes H_Z & 0 
    \end{array} \right), \\
    G_Z &= \left( \begin{array}{cc}
          \vec{a}_Z \otimes \mathbbm{1}_n & 1_{m_Z^C} \otimes \mathbf{H}_Z^T  \\
         \vec{d}_Z^T \otimes H_X & 0 
    \end{array} \right), 
\end{align}
where $m_X^{C}$ ($m_Z^{C}$) denotes the number of rows of $\vec{a}_X$ ($\vec{a}_Z$), the detector matrices are given by
\begin{align}
    D_X &= \left( \mathbbm{1}_{m_X^B} \otimes H_X \quad \vec{a}_X^T \otimes \mathbbm{1}_{r_X} \right), \\
    D_Z &= \left( \mathbbm{1}_{m_Z^B} \otimes H_Z \quad \vec{a}_Z^T \otimes \mathbbm{1}_{r_Z} \right),
\end{align}
and logical operators as
\begin{align}
    L_X = (\vec{g}_X \otimes \ell_X), \\
    L_Z = (\vec{g}_Z \otimes \ell_Z),
\end{align}
Here, the quantities $\vec{a}_X$ and $\vec{d}_X$ are specific to the concrete circuit, however, they already have large similarity with matrices obtained from the hypergraph product of a length-2 chain complex with a length-1 complex.

More evidence that our formalism can also describe fault-tolerant error correction is found by considering the explicit example of the $r$-times repeated measurement of stabilizer generators.
This example is discussed in Sec. VII C of Ref.~\cite{li_low-density_2025}.
For this example, translating the notation of Li to our notation, $\vec{a}_X$ is the check matrix of the repetition code and $g_X$ is the all-one vector, equivalent to what we have obtained in \secref{sec:derivations} from the Künneth formula.
The matrix $\vec{d}_X$ is equivalent to the $r$-dimensional identity matrix $\mathbbm{1}_r$ up to an additional all-zero row.
Similar expressions are obtained when expressing $D_Z$ and $L_Z$.

\section{``Foliation'' beyond the repetition code}

We consider as an example a classical linear code that extends the repetition code.  
In particular, we focus on a code family defined by the polynomial $t(x) = 1 + x + x^2$.  
Note that, in a similar notation, the ordinary repetition code is defined by the polynomial $1 + x$.  

The polynomial $t(x)$ defines a family of classical linear codes with parameters $[n = 3s, k=2, d=2s]$, where the parameter $s > 0$ determines the size $3s \times 3s$ of the circulant matrices used in the code construction.  

Codewords of the classical code take the form
\begin{align}
    \ell_1 &= \underbrace{(1, \dots, 1)}_{s \text{ times}} \otimes (1, 1, 0), \\
    \ell_2 &= \underbrace{(1, \dots, 1)}_{s \text{ times}} \otimes (1, 0, 1). 
\end{align}

Compared to the ordinary repetition code, this code encodes two logical bits, at the cost of a reduced distance.  

We compute the (co-)homology of the chain complex whose boundary map is associated with the check matrix $\tilde{\mathcal{R}}$ of this code.  
The resulting homology groups are
\begin{align}
    H_0(\tilde{\mathcal{R}}) &= \tilde{R}_0 / \mathrm{im}(\tilde{R}) \cong \{ \mathbf{0}, (1, 0, \dots, 0), (0, 1, 0, \dots, 0) \}, \\
    H_1(\tilde{\mathcal{R}}) &\cong \ker(\tilde{R}) \cong \{ \mathbf{0}, \ell_1, \ell_2 \}, \\
    H^0(\tilde{\mathcal{R}}) &\cong \ker(\tilde{R}^T) \cong \{ \mathbf{0}, \ell_1, \ell_2 \}, \\
    H^1(\tilde{\mathcal{R}}) &= \tilde{R}_1 / \mathrm{im}(\tilde{R}^T) \cong \{ \mathbf{0}, (1, 0, \dots, 0), (0, 1, 0, \dots, 0) \}.
\end{align}

As a result, the number of logical correlations and logical errors in the fault complex is increased.  
To be explicit, assuming for simplicity that the base CSS code has full-rank check matrices, the primal logical errors $H_i(\mathcal{F})$ and the primal logical correlations $H^i(\mathcal{F})$ are given by
\begin{equation} \label{eq:ex_primal-log-error}
\begin{split}
    H_i(\mathcal{F}) &\cong H_0(\tilde{\mathcal{R}}) \otimes H_i(\mathcal{C}) \oplus H_1(\tilde{\mathcal{R}}) \otimes H_{i-1}(\mathcal{C}) \\
    &= \langle ( (1, 0, \ldots, 0) \otimes \ell_Z, (0, 1, 0, \ldots, 0) \otimes \ell_Z, \mathbf{0} ) \rangle, 
\end{split}
\end{equation}
where $\ell_Z \in \ker(\partial_i) / \mathrm{im}(\partial_{i+1})$.  

Similarly, the primal logical correlations are
\begin{equation} \label{eq:ex_primal-log-corr}
\begin{split}
    H^i(\mathcal{F}) &\cong H^0(\tilde{\mathcal{R}}) \otimes H^i(\mathcal{C}) \oplus H^1(\tilde{\mathcal{R}}) \otimes H^{i-1}(\mathcal{C}) \\
    &= \langle ( \ell_1 \otimes \ell_X, \ell_2 \otimes \ell_X, \mathbf{0} ) \rangle,
\end{split}
\end{equation}
where $\ell_X \in \ker(\partial_{i+1}^T) / \mathrm{im}(\partial_i^T)$.  

We emphasize that, due to the structure of the check matrix $\tilde{R}$—as evident from its polynomial description $t(x) = 1 + x + x^2$—each bit is involved in three checks.  
As a consequence, the space-time code described by $\mathcal{F}$ is non-matchable, even if the underlying (static) CSS code chain complex $\mathcal{C}$ is associated with a matchable code.  

Additionally, we note that, while the graph state / fault complex now hosts an additional logical correlation and logical error, it is not immediately obvious how to operate on this correlation.  
A similar issue arises in constructing CSS codes with multiple logical qubits.  

Nevertheless, it may be beneficial to consider codes beyond the repetition code for ``foliation'' when the base code is already a non-matchable LDPC code.  
In such cases, the decoding problem already requires algorithms such as BP-OSD, making this extension a natural choice.

\end{document}